# REDUCING BUILDING HEAT DEMAND THROUGH INTELLIGENT CONTROL: A COMPARATIVE SIMULATION STUDY

**Schilt, Ueli[1,2]; Meister, Curtis[1]; Schuetz, Philipp[1]**

1: Competence Centre for Thermal Energy Storage.
School of Engineering and Architecture, Lucerne University of Applied Sciences, Horw, Switzerland.
e-mail: ueli.schilt@hslu.ch; curtis.meister@hslu.ch; philipp.schuetz@hslu.ch;

2: School of Architecture, Civil and Environmental Engineering.
École Polytechnique Fédérale de Lausanne (EPFL), Lausanne, Switzerland.
e-mail: ueli.schilt@epfl.com

## ABSTRACT

Space heating remains the dominant energy consumption in buildings. While structural retrofitting is an effective means to reduce demand, it is often costly and time-intensive. As an alternative, this study examines the potential of intelligent heating control strategies to reduce heat consumption with lower investment and faster implementation. Previous studies have shown that replacing conventional heating-curve-based controllers with intelligent model predictive controllers (MPCs) can reduce heating energy demand. Whereas many studies quantify the benefits of MPC relative to conventional control, this work compares two MPCs with different control objectives in a simulation study. The goal is to quantify the effect of control strategy choice on both indoor temperature tracking and heat demand of a building. We develop a virtual building model in Python, based on ISO 52016-1, to generate high-resolution synthetic measurement data for a residential building. A simple resistance-capacitance (RC) model is parametrised using this dataset and serves as the internal model for two distinct MPC strategies. Both strategies are implemented in MATLAB using the MPC Toolbox and differ only in their optimisation objective: one minimises quadratic heating power at each time step, while the other prioritises temperature tracking for improved thermal comfort. In most building applications, MPC is designed to minimise cost, energy use, or emissions, and it is uncommon for it to be applied solely for setpoint tracking, making this approach a novel aspect of the study. Simulation results over six days show that both strategies meet comfort and system constraints, but their performance differs in terms of energy use and temperature variation. The controller optimising for thermal comfort achieves a lower total heat consumption than the one minimising heating power, which is attributed to the penalisation of high heating power rates in the quadratic cost function. These results demonstrate the influence of objective function formulation on system performance and suggest that MPC-based control can achieve high comfort levels with comparable or reduced heating demand, even without structural changes to the building envelope.

## 1. INTRODUCTION

Buildings account for approximately 40% of global final energy consumption, with a large portion attributable to space heating and cooling [1]. Hydronic heating systems have traditionally been regulated by simple weather-compensation controllers (heating curve controllers) that adjust the heat supply based solely on the current outdoor ambient temperature and rely on simple thermostatic valves to prevent overheating. While easy to implement, these reactive ambient-temperature-only controllers often lack more sophisticated feedback from indoor conditions and usage patterns, which can lead to inefficient overheating of interior spaces and overconsumption of energy when solar or internal heat gains are not accounted for [2]. Improving upon such basic control strategies is critical for avoiding comfort issues and excessive energy use in buildings.

In recent years, advanced control approaches like model predictive control (MPC) have shown great promise for improving building heating efficiency [3], [4]. MPC uses a predictive model of the building's thermal dynamics along with forecasts (e.g. weather and occupancy) to optimise heating control decisions over a receding time horizon. At each time step, MPC solves an optimisation problem that minimises a defined cost function (such as energy use or operation cost) subject to comfort constraints and then applies the first optimal control action before repeating the process at the next interval. By anticipating future demand and avoiding unnecessary heating, MPC controllers can reduce heating energy consumption compared to conventional rule-based or thermostatic controllers. For instance, a recent field study in a residential building demonstrated that a predictive control strategy achieved between 26% and 49% savings in combined heating and cooling energy relative to the baseline hysteresis controller [5]. Numerous simulation and real-world studies similarly report double-digit percentage reductions in heating demand when using MPC instead of traditional controls [6], [7], [8], [9].

A key aspect of any MPC strategy is its control objective, which dictates the trade-off between comfort and energy use [10]. In building climate control, MPC is often formulated to minimise heating energy consumption while maintaining indoor temperatures within acceptable comfort ranges [11]. However, different MPC implementations may prioritise objectives differently depending on the goals [12]. Some controllers are tuned to track the desired indoor temperature setpoint very tightly (maximising comfort), whereas others adopt an economic target that prioritises reducing energy or cost, allowing slight deviations from the setpoint as long as comfort bounds are respected [13]. The choice of objective function can significantly affect performance [14]. For example, an MPC scheme that aggressively eliminates any temperature deviations (focusing purely on setpoint tracking) can increase heating energy use [11]. On the other hand, an MPC that emphasises energy minimisation might incur larger deviations from the setpoint temperature to yield lower energy demand, striking a different balance between comfort and efficiency [15]. Each formulation, therefore, entails a distinct comfort–energy trade-off.

While previous studies commonly compare conventional controllers with MPCs, this work focuses on differences between two MPC control objectives. The goal is to evaluate how these control strategies compare in terms of temperature tracking (i.e. thermal comfort) and building heat demand. Accordingly, in this paper we assess two MPC control strategies for a residential building heating system, each with a different objective. The first strategy (comfort-tracking MPC) is designed to maintain indoor temperature as close as possible to the setpoint, prioritising thermal comfort. In most building control applications, MPC is formulated to minimise cost, energy use, or emissions, and it is uncommon to employ it solely for setpoint tracking, making this a novel focus in the present study. The second strategy (power-minimising MPC) aims to minimise heating power usage, focusing on a more constant consumption while keeping temperature within acceptable comfort limits. We perform simulations using both approaches and evaluate their outcomes in terms of the achieved comfort level (temperature regulation and any deviations) and the resulting heating energy demand. By analysing these two scenarios side-by-side, we not only illustrate how the MPC objective choice influences building comfort and energy performance but also focus on the relevance of objective function formulation. The results provide insights into the potential benefits and trade-offs of comfort-driven versus power-driven predictive control.

## 2. METHODOLOGY

### 2.1. Virtual building

When applying model predictive control (MPC) to a building heating system, the controller relies on a mathematical model of the building's thermal dynamics, referred to as the "internal plant model". This model is used to predict the building's thermal response over a future time horizon and determine the corresponding optimal heating power. In real buildings, the MPC controller receives feedback from sensor measurements (e.g., indoor temperature), which is used to update predictions and improve control accuracy. In simulation studies, such measurements are not available, so a separate mathematical model is created to represent the virtual building. This model, known as the "plant", generates synthetic measurement data and is typically more detailed than the internal plant model used by the MPC.
In this study, we use the procedure described in the ISO 52016-1:2017 standard [16], [17] to implement a building thermal model to be used as plant and generate synthetic measurement data. The ISO standard provides a detailed calculation method for assessing the energy performance of buildings by determining the hourly or monthly energy needs for heating and cooling. It covers internal temperatures and heat loads based on building characteristics, weather data, and occupancy profiles. The standard supports both design and energy performance assessments, and it allows for dynamic (hourly) and quasi-steady-state (monthly) calculations, depending on the desired level of detail. We implement the model in Python programming language and adapt it to generate simulation data in 10-minute resolution. We call it the "ISO model".
First, we generate a set of virtual measurement data simulating the building operation over three months (January to March 2020) for a virtual building located in Zurich, Switzerland, consisting of 100 m$^2$ footprint area and two stories. Figure 1 shows the resulting data for the month of March. Weather data (ambient temperature, solar irradiation) is obtained from Meteostat [18] for the selected timeframe. The heating ($h$) power delivered to the building is calculated at each time step using a simple linear controller that only depends on the ambient ($amb$) temperature:

$$\dot{Q}_h(T_{amb}) = \begin{cases} \dot{Q}_{h,0} - mT_{amb} & \forall\, T_{amb} < T_{cutoff} \\ 0 & \forall\, T_{amb} \geq T_{cutoff} \end{cases} \quad (1)$$

The heating rate $\dot{Q}_h(T_{amb})$ is determined by the ambient temperature $T_{amb}$, the specified heating rate at 0°C ambient temperature $\dot{Q}_{h,0}$, and the heating curve gradient calculated as $m = \frac{\dot{Q}_{h,0}}{T_{cutoff}}$. The cut-off temperature $T_{cutoff}$ represents the ambient temperature above which no active heating is required and therefore the heating is switched off. In this study, we set $T_{cutoff}$ to 12°C.
The simulation results are used to parametrise the internal controller model.

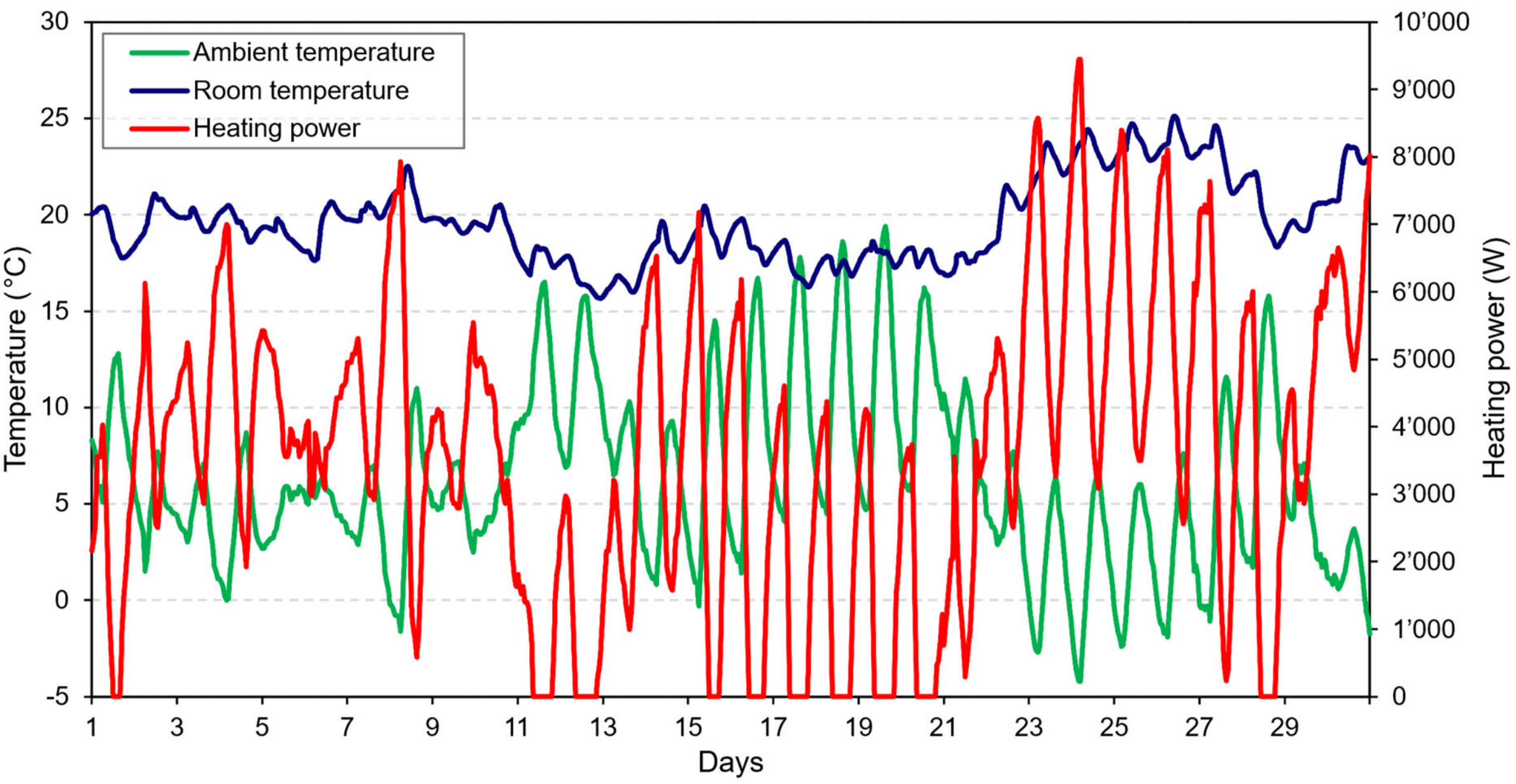

Figure 1: 31 days of simulated indoor temperate using the ISO 52016 building performance model for the month of March 2020. A simple ambient temperature-based linear controller is used to determine the heating power.

## 2.2. Controller model

As an internal plant model for the MPC controller, a Resistance-Capacitance (RC) model is used [19]. RC models fall into the category of grey-box models and can be of varying complexity [20]. In this study, we use a simple single-node 1-R-1-C model (see Figure 2), which is formulated in Equation 2. The model is implemented in MATLAB.

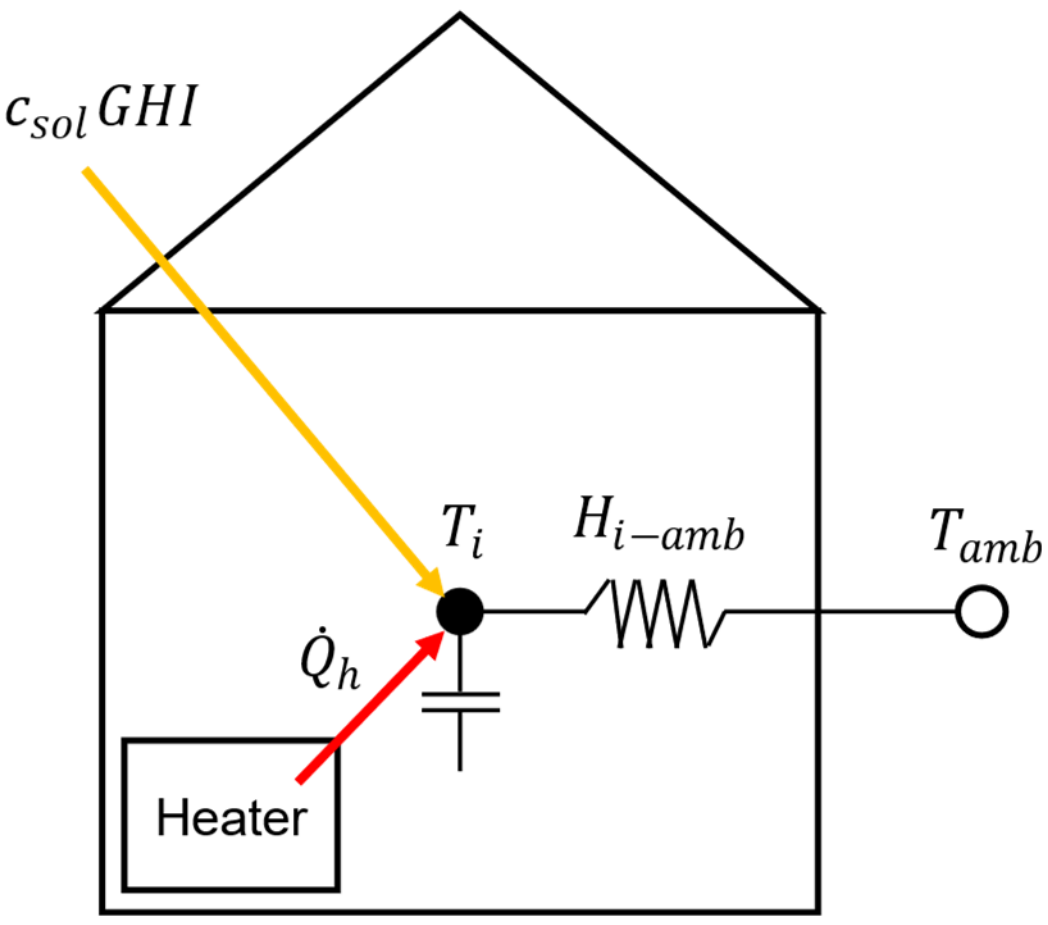

Figure 2: 1-R-1-C model representation of the building's thermal dynamics.

$$C_i \frac{\partial T_i}{\partial t} = H_{i-amb}(T_{amb} - T_i) + \dot{Q}_h + c_{sol}GHI \tag{2}$$

The building's thermal capacity $C_i$ related to interior ($i$) temperature $T_i$, the heat loss coefficient $H_{i-amb}$ from interior to ambient, and the solar gain parameter $c_{sol}$ are determined for the virtual building using the synthetic measurement data generated with the ISO model (see Figure 3). $GHI$ refers to Global Horizontal Irradiation. The parametrisation is carried out with an optimisation minimising the root-mean squared error between synthetic measured room temperature and room temperature predicted by the RC model. Given the parametrisation with the ISO model data, the RC model can now be used to control the virtual building (i.e. the ISO model).

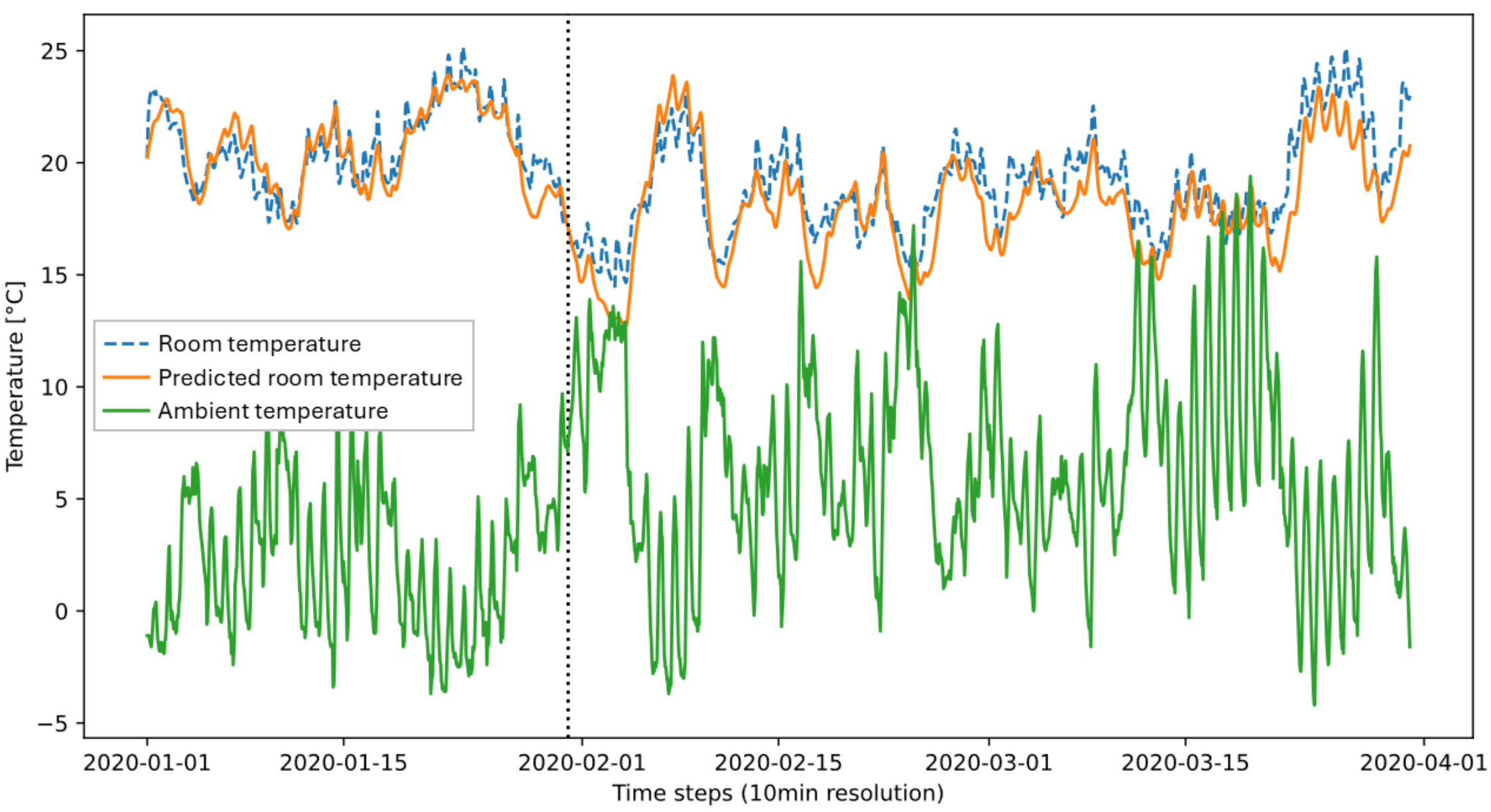

Figure 3: Results of RC model parametrisation using synthetic measurement data over 3 months from the ISO model. Model parameters are determined during an optimisation minimising the root-mean squared error between measured and predicted room temperature.

### 2.3. Model predictive control simulation

To carry out the MPC simulation, we use the MATLAB MPC Toolbox [21]. Since the RC model is implemented in MATLAB, it can directly be integrated in the framework. The ISO model, which is implemented in Python programming language, is incorporated into the MPC framework using a MATLAB-Python interface module.
At each time step, the controller uses the internal plant model to compute the optimal control moves to minimise the objective function over the prediction horizon $H_P$. For this purpose, the controller can decide on the control moves for the intervals contained in the control horizon $H_C$. The optimal control move for the next interval is then applied to the ISO model, and the controller receives feedback from the ISO model to re-compute the optimal control moves in the next step. In this study, the controlled variable is the heating power $\dot{Q}_h$ delivered to the building at each time step and the output is the room temperature $T_i$.
The objective function computes a virtual cost $J$ that arises from deviations from reference values. The optimisation during the MPC operation tries to minimise these costs. Our simulation makes use of a quadratic objective function as provided by the MATLAB MPC Toolbox. A simplified form of the function is shown in equations 3 to 5, which are adapted to the problem in this study, only showing the relevant terms. It consists of two cost terms relating to room temperature ($J_{T_i}$) and heating power ($J_{\dot{Q}_h}$).

$$J(\dot{Q}_h) = J_{T_i}(\dot{Q}_h) + J_{\dot{Q}_h}(\dot{Q}_h) + J_{\Delta\dot{Q}_h}(\dot{Q}_h) + J_{\epsilon}(\dot{Q}_h) \quad (3)$$

$$J_{T_i}(\dot{Q}_h) = \sum_{ts=1}^{p} \left[ \frac{w_{T_i}}{s_{T_i}} (T_{i,ts} - T_{ref,ts}) \right]^2 \quad (4)$$

$$J_{\dot{Q}_h}(\dot{Q}_h) = \sum_{ts=0}^{p-1} \left[ \frac{w_{\dot{Q}_h}}{s_{\dot{Q}_h}} (\dot{Q}_{h,ts} - \dot{Q}_{h,ref,ts}) \right]^2 \quad (5)$$

In equations 4 and 5, parameter $p$ is the number of intervals contained in the prediction horizon, subscript $ts$ is the time step, parameter $s$ is a scale factor in the respective units, and $w$ is a unitless weight attributed to the specific objective. The values of the weights determine which objective (heating power versus temperature tracking) is prioritised in the optimisation. $\dot{Q}_{h,ref,ts}$ is a reference value for the heating power, which is set to 0 in order to minimise energy consumption.
In addition, the following constraints can be set: Minimum and maximum heating power ($\dot{Q}_{h,min}$ and $\dot{Q}_{h,max}$), minimum and maximum heating power difference from one time step to the next ($\Delta\dot{Q}_{h,min}$ and $\Delta\dot{Q}_{h,max}$), and lower and upper room temperature limits ($T_{i,ll}$ and $T_{i,ul}$).
In this study, we run two simulations, A and B. Simulation A focuses on the objective of minimising the quadratic heating power ($w_{\dot{Q}_h} = 1, w_{T_i} = 0$), while simulation B prioritises the objective of tracking the reference room temperature ($w_{\dot{Q}_h} = 0, w_{T_i} = 1$). The simulations are run using 10-minute time intervals. In both scenarios, the reference room temperature is set to 20.0°C. A night-time reduction is also implemented, reducing the reference temperature to 18.0°C between 22:00 and 06:00. Both prediction and control horizon are set to 12 hours. The minimum required heating power is 0 kW, and the maximum allowed heating power is 10 kW. The maximum heating power rate of change is 10 kW in either positive or negative direction. The lower and upper limits of the reference room temperature are set to $T_{i,ll} = T_{ref} - 0.5$°C and $T_{i,ul} = T_{ref} + 5.0$°C, respectively. The simulations are both run for a timeframe of 144 hours (i.e. 6 days).

## 3. RESULTS

Apart from differing objectives, both simulations (A and B) are subject to the same input parameters and constraints. The results differ between the two simulations in terms of heating energy consumed, maximum heating power, and achieved room temperature deviations (Table 1). Simulation A tries to minimise the heating power and results in a maximum value of 7.1 kW, as opposed to simulation B where a maximum heating power of 10.0 kW is observed several times over the 6 days. However, Simulation A has a slightly higher overall heat energy demand of 603 kWh compared to Simulation B with 590 kWh. Accordingly, Simulation A shows a slightly higher mean room temperature (19.54°C) compared to Simulation B (19.41°C). The deviations from the reference room temperature are larger in Simulation A, resulting in a root-mean squared error (RMSE) of 0.82°C, compared to 0.30°C in Simulation B. Both simulations neither violate the heating power constraints nor the upper or lower room temperature constraint. The inputs and outputs at each time step over 6 days are plotted for simulations A and B in Figure 4 and Figure 5, respectively.

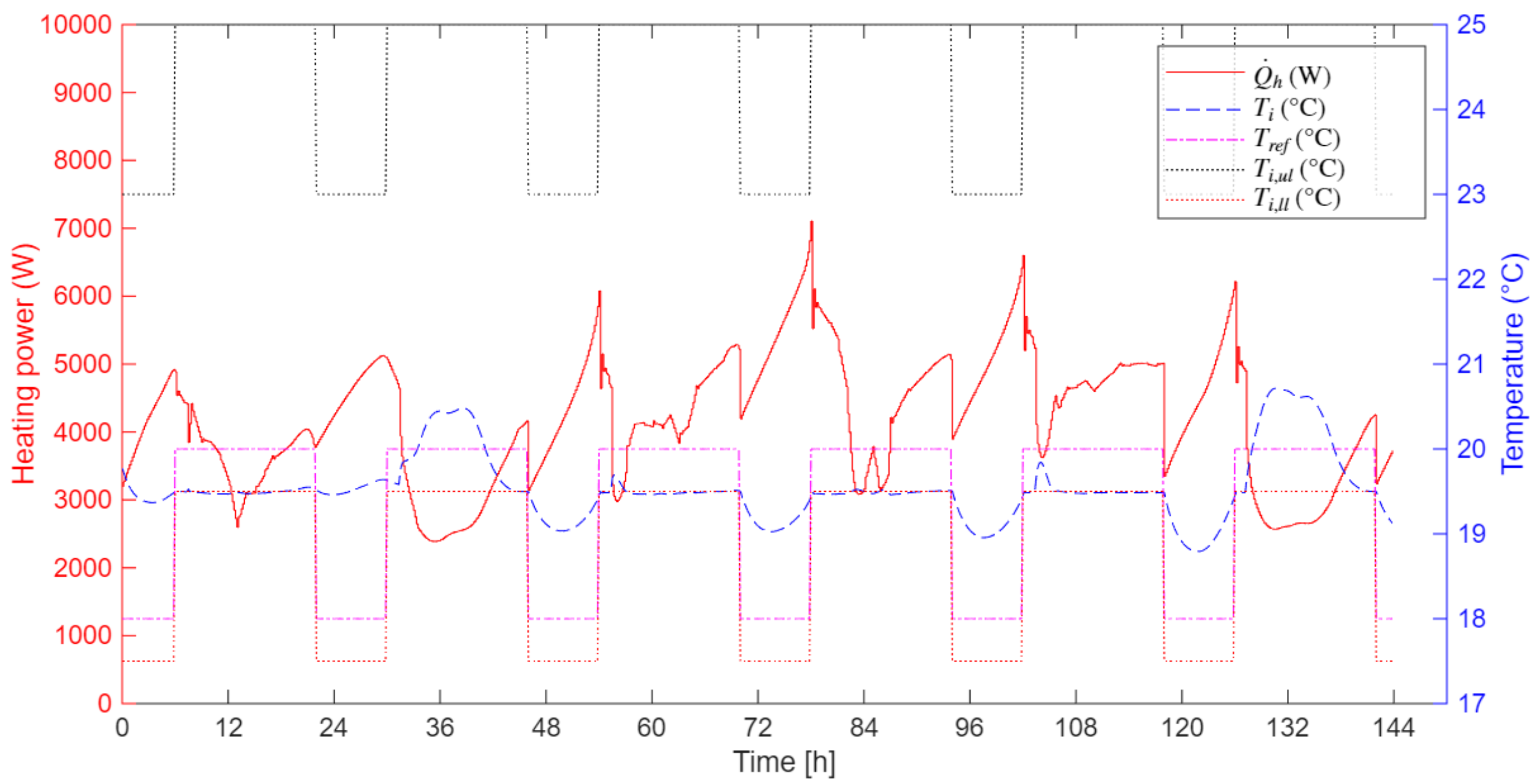

Figure 4: Simulation A – measured input and output. The controller is optimising for minimised heating power. Shown are the heating power ($\dot{Q}_h$), resulting room temperature ($T_i$), reference room temperature ($T_{ref}$), and lower and upper room temperature constraints ($T_{i,ll}$ and $T_{i,ul}$).

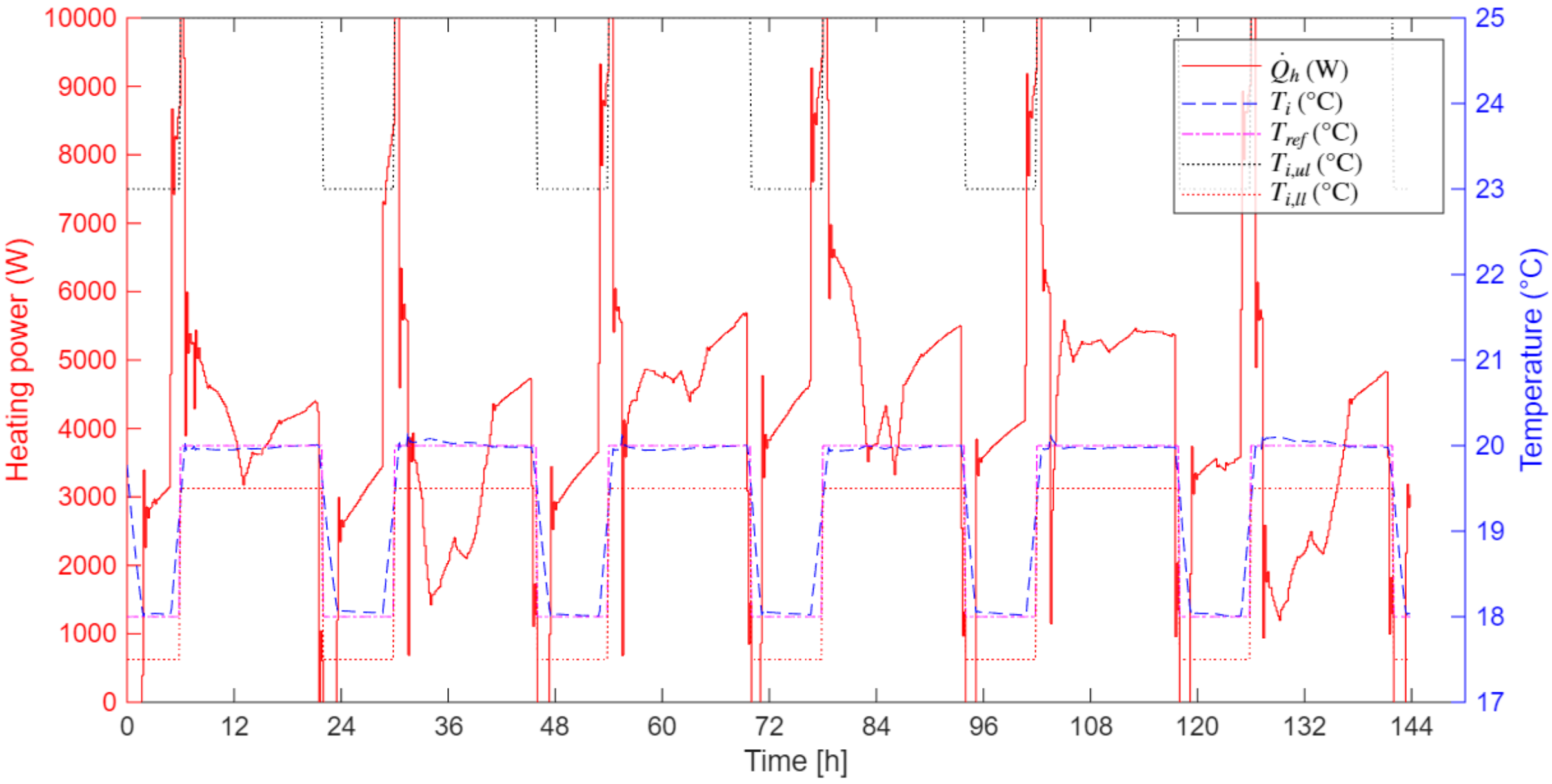

Figure 5: Simulation B – measured input and output. The controller is optimising for temperature tracking (i.e. thermal comfort). Shown are the heating power ($\dot{Q}_h$), resulting room temperature ($T_i$), reference room temperature ($T_{ref}$), and lower and upper room temperature constraints ($T_{i,ll}$ and $T_{i,ul}$).

Table 1: Resulting metrics of simulation A (heating power minimisation) and B (temperature tracking): maximum measured heating power ($\dot{Q}_{h,max,meas}$), cumulative heating energy, mean achieved room temperature ($T_{i,mean}$), and root-mean squared error (RMSE) between reference and achieved room temperature.

| Sim. | Optimisation objective | $\dot{Q}_{h,max,meas}$ | Cumulative heating energy | $T_{i,mean}$ | RMSE |
|---|---|---|---|---|---|
| A | Minimise heating power | 7’107 W | 603 kWh | 19.54°C | 0.82°C |
| B | Temperature tracking | 10’000 W | 590 kWh | 19.41°C | 0.30°C |

## 4. DISCUSSION AND CONCLUSIONS

In this study, we compare two MPC controllers with different optimisation objectives in a simulation study: one optimising for minimised heating power (A) and the other for thermal comfort (B). It is observed that all the defined constraints (heating power limits, temperature limits) can be met under the given ambient conditions. The controller optimising for thermal comfort performs better at tracking the specified reference temperature, resulting in a lower RMSE. This result is expected, as it is the controller's main objective. On the other hand, it would intuitively be expected that the controller optimising for minimised heating power also reaches a lower total heat demand. This, however, is not the case: instead, the controller uses more total heating energy. It is observed that this controller uses less extreme heating power values at individual time steps, never using the maximum allowed input of 10 kW. However, the total energy supply is 2% larger. One possible explanation for this behaviour is the quadratic nature of the objective function: In Equation 5, large heating power values at individual time steps lead to a disproportionately high cost $J_{\dot{Q}_h}$ because its value is squared. Therefore, the controller avoids extreme values at individual time intervals. This strategy, however, leads to a larger overall energy consumption. The controller optimising for thermal comfort, on the other hand, regularly utilizes very large heating power values at individual time steps and this strategy seems to lead to a slightly lower energy consumption. This leads to two conclusions: Firstly, the deployment of MPC can achieve a high level of thermal comfort, possibly without compromising much on energy demand. Secondly, it demonstrates the importance of objective function formulation: A different formulation (e.g. non-quadratic objectives or direct energy minimisation instead of power minimisation) can focus on the overall energy demand and might be able to reduce the energy consumption further. This remains to be investigated in future studies.
It should be noted that the results and conclusions presented here are specific to the simulated virtual building and the ambient conditions used in this study. Different building characteristics or climatic conditions could lead to different behaviours. Future work could therefore extend the analysis to a broader range of building types and climates, and to real-world MPC implementations, in order to validate the findings beyond simulation.

## 5. ACKNOWLEDGEMENTS

The research presented in this article was supported by InnoSuisse (grant:101.224 IP-EE) and by the Swiss Federal Office of Energy as part of the SWEET EDGE and PATHFNDR consortium. In this context, the authors bear sole responsibility for the conclusions and the results presented in this publication.